\documentclass[preprint]{aastex}

\slugcomment{}

\shorttitle{Second generation stars in GCs}
\shortauthors{Carretta et al.}

\begin{document}


\title{Properties of second generation stars in Globular Clusters\altaffilmark{1}}


\author{Eugenio Carretta\altaffilmark{2}, Angela Bragaglia\altaffilmark{2}, 
Raffaele G. Gratton\altaffilmark{3}, Sara Lucatello\altaffilmark{3} }

\altaffiltext{1}{Written version of a contributed talk presented at the
conference ``Chemical Evolution of Dwarf Galaxies and Stellar Clusters",
Garching bei M\"unchen, Germany, July 21-25, 2008}
\altaffiltext{2}{INAF-Osservatorio Astronomico di Bologna, via Ranzani 1, 
I-40127 Bologna, Italy}
\altaffiltext{3}{INAF-Osservatorio di Padova, Vicolo dell'Osservatorio 5, 
I-35122 Padova,Italy}


\begin{abstract}

We present the first results from the analysis of GIRAFFE spectra of more than
1200 red giants stars in 19 Galactic Globular Clusters (GCs), to study the
chemical composition of second generation stars and their link with global
cluster parameters. We confirm that the extension of the Na-O anticorrelation
(the most striking signature of polluted, second generation populations) is
strictly related to the very blue (and hot) extreme of the Horizontal Branch
(HB). Long anticorrelations seem to  require large mass and large-sized,
eccentric orbits, taking the GCs far away from the central regions of the
Galaxy.  We can separate three populations in each cluster (primordial,
intermediate and extreme) based on the chemical composition. In all GCs we
observe a population of primordial composition, similar to field stars of
similar metallicity. We find that in all GCs the bulk (from 50 to 70\%) of stars
belong to the intermediate component. Finally, the extreme, very oxygen-poor
component is observed preferentially in massive clusters, but is not present in
$all$ massive GCs.

\end{abstract}


\section{Introduction}

\begin{figure*}[h]
\centering
\includegraphics[scale=0.85]{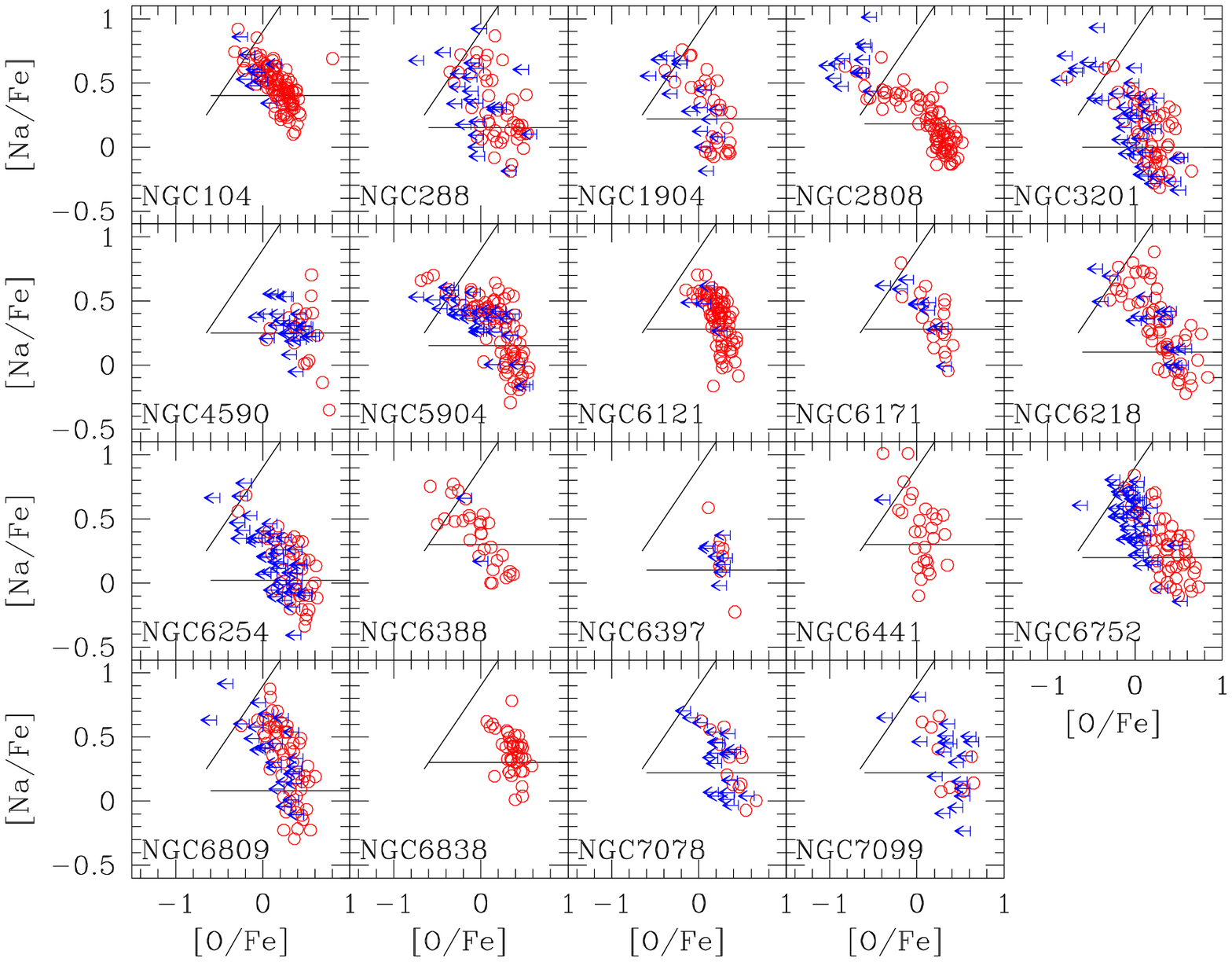}
\caption{Summary of the Na-O anticorrelation observed in the 19 globular
clusters of our sample. Arrows indicate upper limits in oxygen abundances. The 3
solid lines in each panel separate the Primordial component (located in the 
Na-poor/O-rich region), the Na-rich/O-poor Extreme component, and the Intermediate component in
between (see text for more details).}
\end{figure*}

Galactic Globular Clusters (GCs) show a series of star-to-star abundance
variations for several light elements. Part of them (Li, C, N) show the same
behavior in field stars, some others (O, Na, Mg, Al) show a behavior peculiar to
GCs. In particular, there are anticorrelations between O and Na, Al and Mg: they
are present in all evolutionary phases sampled (both unevolved main sequence and
evolved red giant branch stars) in all GCs studied so far (for a recent review
see Gratton, Sneden \& Carretta 2004).

The fact that these anticorrelations are present also in unevolved stars
(Gratton et al. 2001, Ramirez \& Cohen 2002, Carretta et al. 2004), unable to
activate the required nucleosynthetic chains (p-capture in H burning at very
high temperatures, e.g. Denisenkov \& Denisenkova 1989) implies that {\em they
have been imprinted in the gas by a previous generation and that we are seing a
second generation of stars in these GCs.} We do not yet know what kind of stars
produced the pollution; the most popular candidates are intermediate-mass AGB
stars (e.g. D'Antona \& Ventura 2007), or very massive, rotating stars
(e.g.Decressin et al. 2007). 

The Na-O anticorrelation is also stricly connected to the helium abundance
(Na-rich, O-poor stars should also be He-rich) and to the multiple sequences
that have been recently found in some GCs ($\omega$ Cen, NGC~2808, see
Bedin et al. 2004 and Piotto et al. 2007, respectively) and
attributed to populations with different Y.

There are currently data on more than 1400 stars, in about 20\% of known
GCs, thanks to
the efforts of many researchers, {\it in primis} the Lick-Texas group, lead by
Chris Sneden and Robert Kraft; our group is now one of the main 
contributors, with more than 1200 stars analyzed. The bottom line is that we
can stop talking about "abundance anomalies" and we must start speaking about
the normal chemical composition of second generation stars in GCs. 
Times are ripe to start a more quantitative approach.

Therefore, with the goal of understanding what is the relation between these
abundance patterns and the horizontal branch (HB) morphology or  structural parameters of the
GCs, we have conducted an homogeneous survey of RGB stars in 19\footnote{
NGC~104, NGC~288, NGC~1904, NGC~2808, NGC~3201, NGC~4590, NGC~5904, NGC~6121,
NGC~6171, NGC~6218, NGC~6254, NGC~6388, NGC~6397, NGC~6441, NGC~6752, 
NGC~6809, NGC~6838, NGC~7078, and NGC~7099} Galactic GCs
using the multiobject spectrograph FLAMES@VLT.
We obtained Giraffe spectra (at R$\simeq$20000, comprising the Na~I 568.2-568.8nm,
615.4-6.0nm and [O I] 630nm lines) of about 100 stars per cluster, and UVES
spectra (at R$\simeq$40000, covering the 480-680nm region) of about 10 to 14 stars per
cluster. 

We have targeted GCs with [Fe/H] ranging from $-2.4$ to $-0.4$ dex, with
very different HB morphology (stubby red HB, blue HB, very extendend blue HB,
bimodal HB), and very different global parameters (total mass, concentration,
age, orbit, etc.). 

We have determined atmospheric parameters  and measured Fe, O, and Na abundances
for more than 2000 stars (of which more than 1200 cluster members and with
$both$ O and Na detected.

In Fig.~1 we show a collage of the Na-O anticorrelations observed in all 19
clusters in our sample. Solid lines separate the Primordial, Intermediate and
Extreme populations, discussed below.

The number of stars measured in each cluster depends on the richness of
population, metallicity, S/N ratio and in some cases on field stars
contamination, as in the case of the bulge clusters NGC~6388 and NGC~6441, 
or the disk cluster NGC~6171.

\section{Extension of the Na-O anticorrelation and (blue) HB}

As a way to quantify the extension of the anticorrelation, we adopted the
interquartile range (the middle 50\% of the data) of the [O/Na] distribution.
This was suggested by Carretta (2006) as the best indicator, because it is not
much sensitive to the ouliers in the distribution. 

\begin{figure}[ht]
\centering
\includegraphics[scale=0.40]{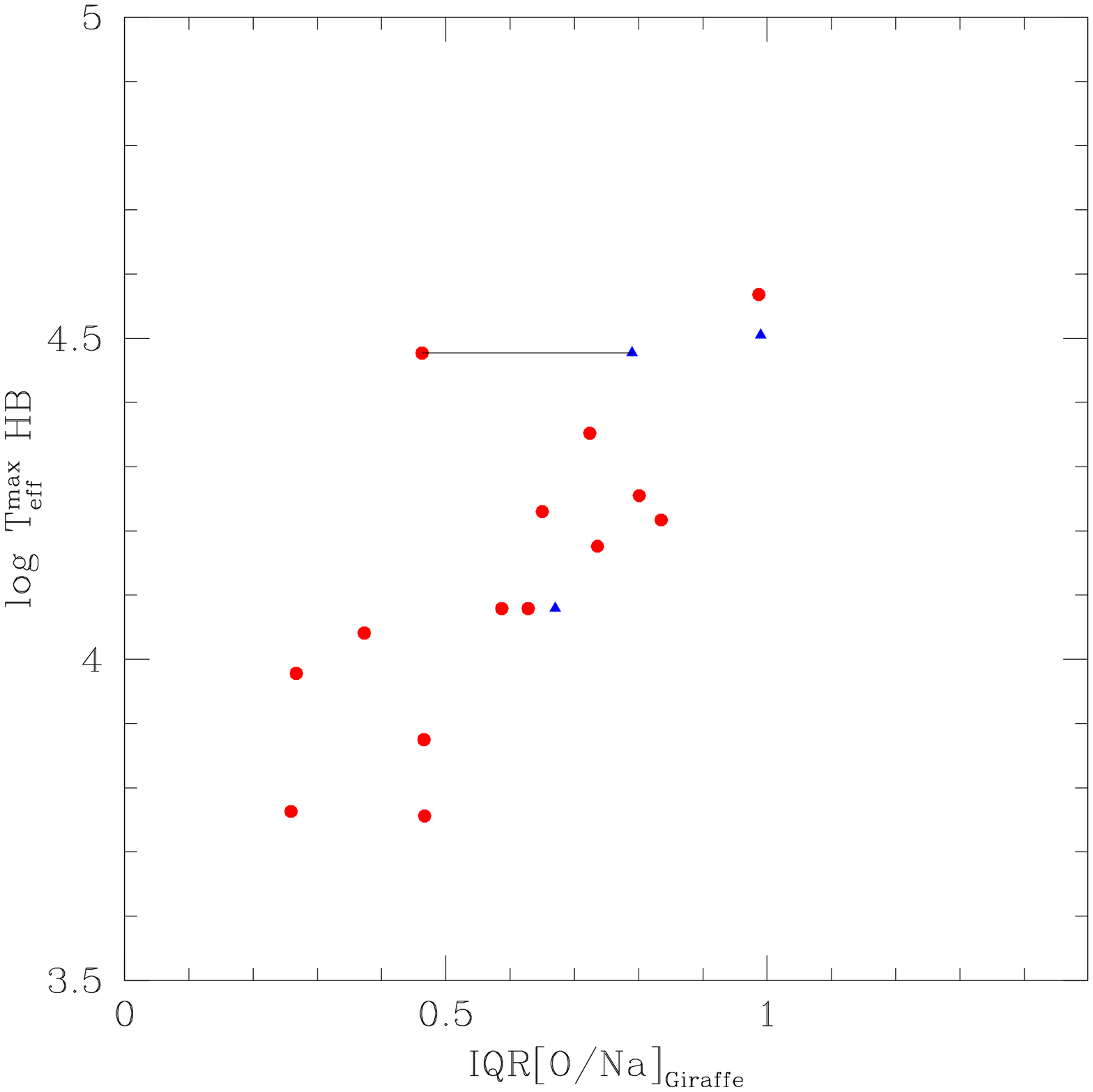}
\caption{Interquartile range of the [O/Na] ratio as a function of the maximum
temperature reached on the HB (taken from Recio-Blanco et al. 2006). Filled red
circles are the GCs of our sample, filled blue triangles are estimates for 3
clusters from the literature. The solid line connects the values for M~15 (see
text).}
\end{figure}

The most tight correlation we found (shown in Fig.~2) is between the extension
of the Na-O anticorrelation (well measured by the Inter-Quartile range
IQR[O/Na]) and of the HB (measured by the maximum Teff reached on the HB). This
has been first demonstrated by Carretta et al. (2007d) on a subsample of our 19
GCs and literature data, and amply confirmed on our whole sample. In Fig.~2 the
red  dots are for our data, the blue triangles for literature data: M~15 (Sneden
et al. 1997),  NGC~362 (Shetrone \& Keane 2000), and M~13 (Sneden et al. 2004,
Cohen \& Melendez 2005). 

M~15 is somewhat uncertain: the value from literature falls very well on the
relation defined by all other clusters, while our value is more
discrepant: maybe we missed some very O-poor stars, difficult to measure at this
very low metallicity.
The Spearman rank coefficient tells us that
the probability to get such a tigh relation only by chance is negligible,
ranging from 1 in 10$^{-3}$ if we include our value for M 15, down to 1 
in 10$^{-5}$ or 4 in 10$^{-6}$ excluding M 15 or taking the Lick value.

We conclude that this is a real, very strong relation: the first conclusion we
can draw is that a same mechanism drives/affects the extent of the pollution on
the RGB and the morphology of the bluest end of the HB.

Notice, however, that this can be only considered $a$ second parameter: 
the global distribution of stars on the HB (as indicated e.g., by the HB 
ratio (B+R)/(B+R+V) -see Harris 1996 and web updates) is $not$
correlated with the IQR[O/Na], i.e. with the extension of the Na-O
anticorrelation.

\section{Relations with structural and orbital parameters}

We also found a good correlation between the IQR[O/Na] and the total mass (or
absolute magnitude M$_V$) of the GCs (see Fig.~3): an high mass seems to be a
requisite for an extended anticorrelation. 

\begin{figure}[ht]
\centering
\includegraphics[scale=0.40]{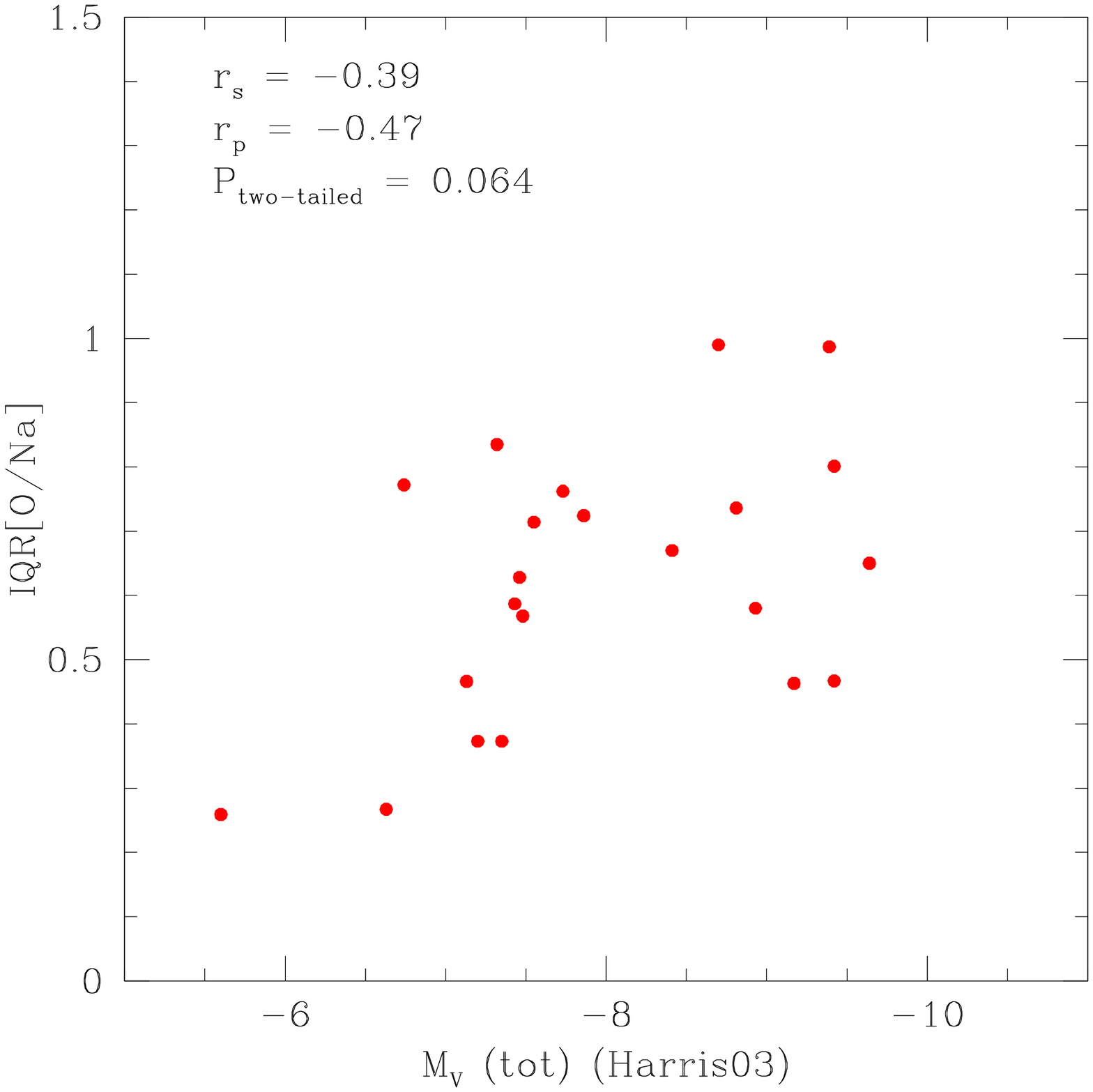}
\caption{The extension of the Na-O anticorrelation (measured using the
IQR[O/Na]) as a function of the clusters' absolute magnitude M$_V$ from Harris
(1996 -as updated in the web page). The Pearson and Spearman correlation
coefficients are listed.}
\label{f:procfig3}
\end{figure}

This may be explained by a better capability of the cluster to retain the gas
polluted by the first-generation stars to form the second-generation stars in
which we measure the anticorrelated Na and O. However, high mass is not a
sufficient condition:  47~Tuc is a notable counter-example. This  cluster simply
does not show very O-poor stars and presents a short/normal Na-O
anticorrelation even if it is a very massive object. Some other factors must be
involved.
On the other hand, we note that the position of GCs standing leftward of the
main relation (namely NGC~288, NGC~6218 and M~71) is more easily explainable
by the fact that they may have lost a (probably
large) fraction of their mass in the past.

We explored the relations with orbital parameters, as first
suggested by Carretta (2006), and performed simple bivariate correlations,
adding to the cluster absolute magnitude a dependence on the cluster orbital
parameters. In Fig.~4 we show the IQR[O/Na] as a function of a combination of
M$_V$ and the eccentricity of the orbits, whenever 
available (Dinescu et al. 1999).

\begin{figure}[ht]
\centering
\includegraphics[scale=0.40]{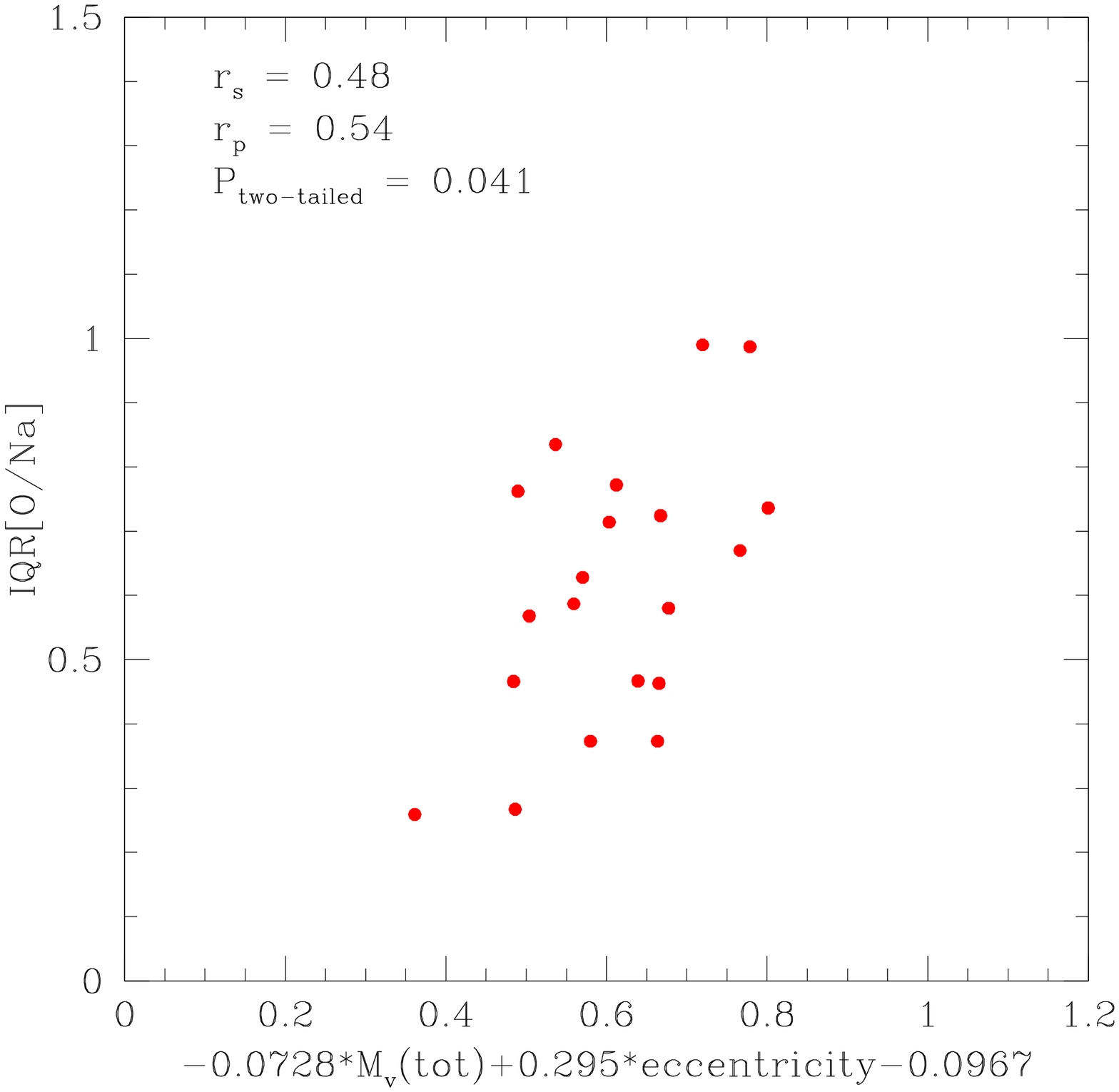}
\caption{The extension of the Na-O anticorrelation (measured using the
IQR[O/Na]) as a function of a combination of the clusters' absolute magnitude 
M$_V$ and of the eccentricity of the orbits in the Galaxy.}
\end{figure}

When we combine mass (magnitude) and orbital parameters 
the relations get tighter and statistically more significant. More
importantly, 
47~Tuc does not stand out at all in this plot, and is now virtually
indiscernible from the other clusters. The same holds using other orbital
parameters (as total energy, or period of the orbit, or maximum height over the
plane).

We consider these results as an indication that also the initial
position and the time spent away from possible perturbations affect the
possibility of a GC to build a gas reservoir from which the second-generation
stars are born.

\section{The fraction of stars with Primordial, Intermediate and Extreme
composition}

The  ratio of first to second-generation stars is very useful to put constraints
on the formation scenario. We assume as first-generation stars (or primordial,
P) those with O and Na similar to field stars of same metallicity.   
We divide the
second-generation stars according to the pollution degree in Intermediate (I)
and Extreme (E: those with [O/Na]$<-0.9$ dex, from a comparison with the
distribution in NGC~2808, see Carretta et al. 2006)

We have applied this operative division to all the 19 GCs in our
sample. The lines separating the three components are shown in Fig.~1.  
From this figure and computing the fraction of stars in each component we found 
that:
\begin{itemize}
\item the Primordial population is present in all GCs.
\item the Intermediate 2nd generation constitutes the bulk (50-70\%) of stars.
\item the Extreme 2nd generation is not present in all GCs.
\end{itemize}

The fraction of Intermediate and Extreme second generation stars are
anticorrelated, and the Intermediate population is anticorrelated with the
IQR([O/Na] (Fig.~5) since, by definition, the extension of the Na-O 
anticorrelation is driven by the Extreme population (Fig.~6).

\begin{figure}[ht]
\centering
\includegraphics[scale=0.40]{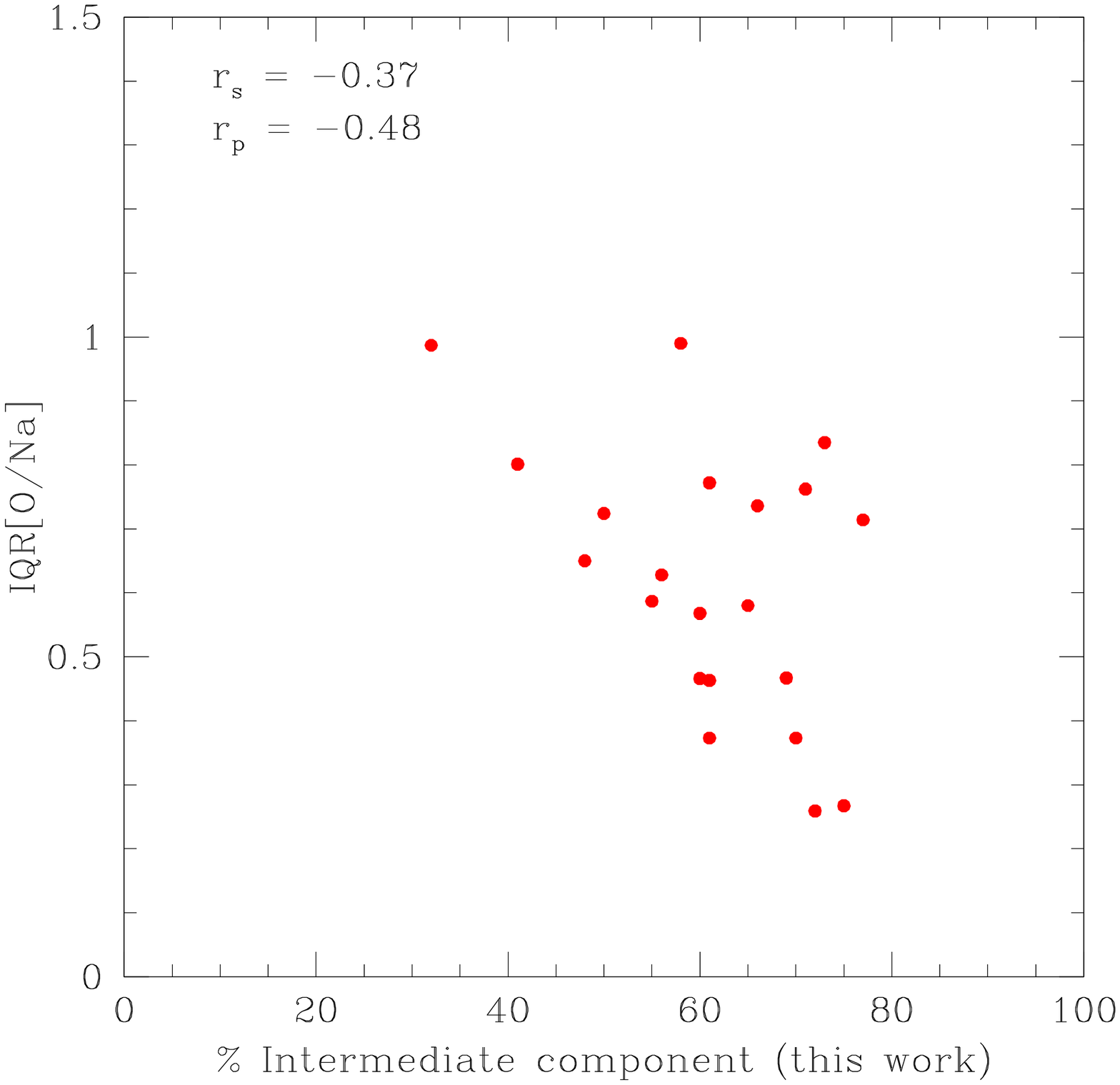}
\caption{The extension of the Na-O anticorrelation as a function of the fraction
of stars in the intermediate I component of second-generation stars.}
\end{figure}

\begin{figure}[ht]
\centering
\includegraphics[scale=0.40]{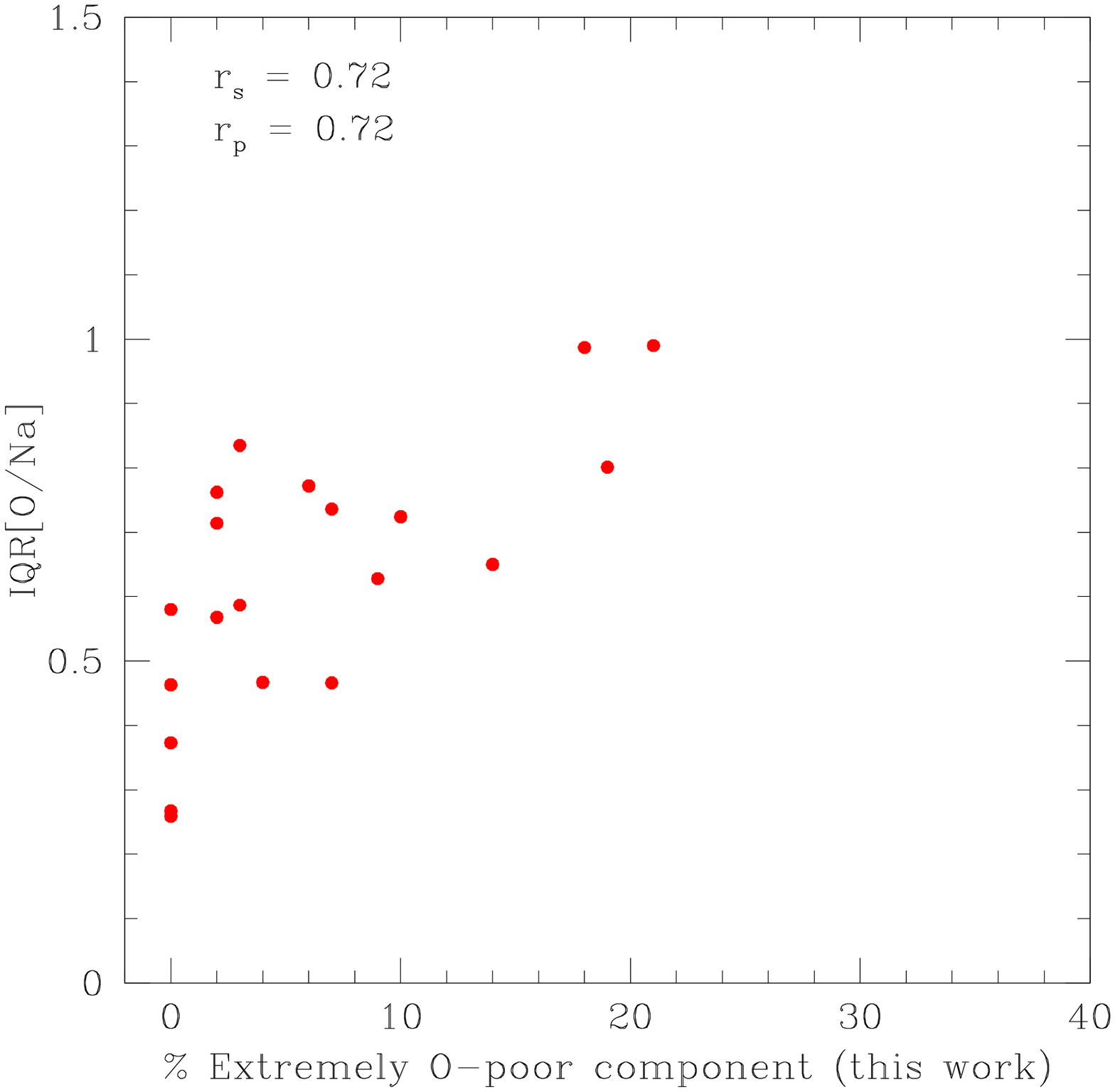}
\caption{The extension of the Na-O anticorrelation as a function of the fraction
of stars in the extreme E component of second-generation stars.}
\end{figure}

When the fractions of the three components are plotted $vs$ total mass (measured
by total M$_V$, Fig.~7), we see that less Primordial stars are lost by the more
massive clusters, as expected. As with the IQR[O/Na], the Extreme population is
more pronounced in heavier GCs, but the relation is not one-to-one. 

\begin{figure}[ht]
\centering
\includegraphics[bb=50 140 290 710, clip, scale=0.44]{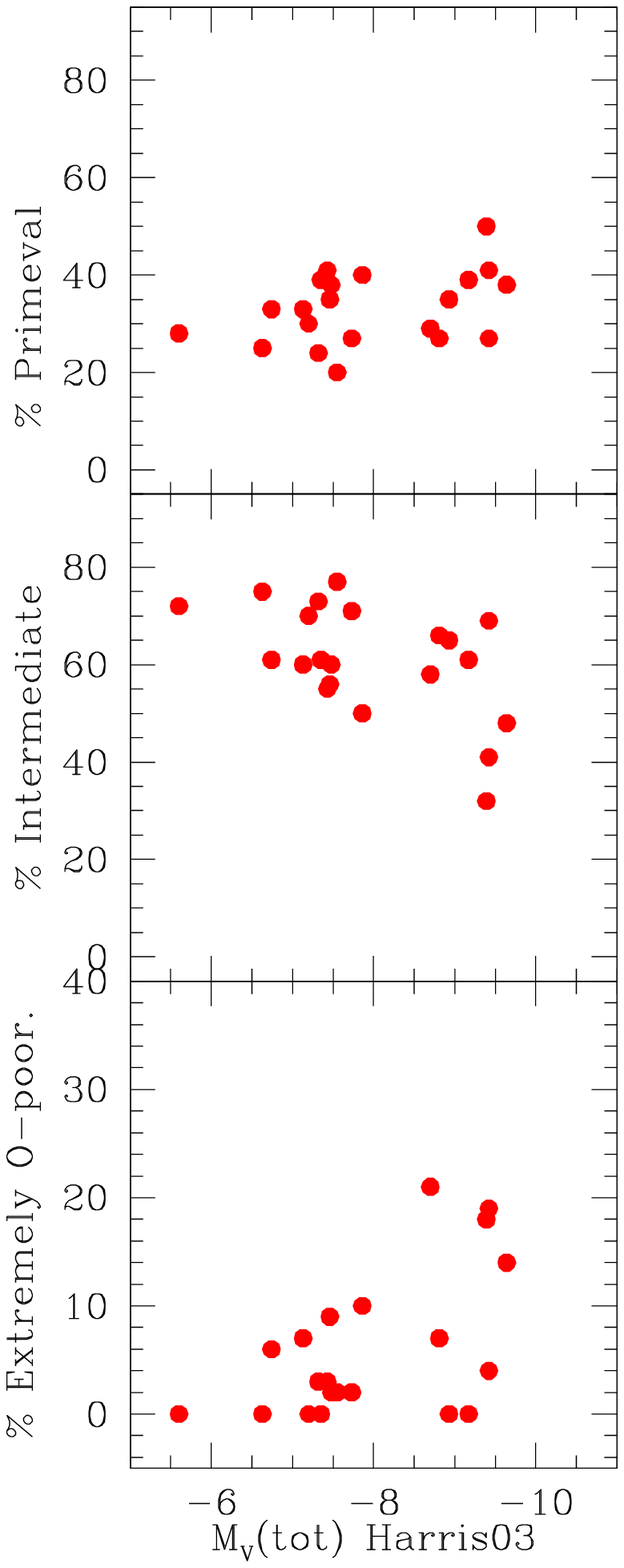}
\includegraphics[bb=50 140 290 710, clip, scale=0.44]{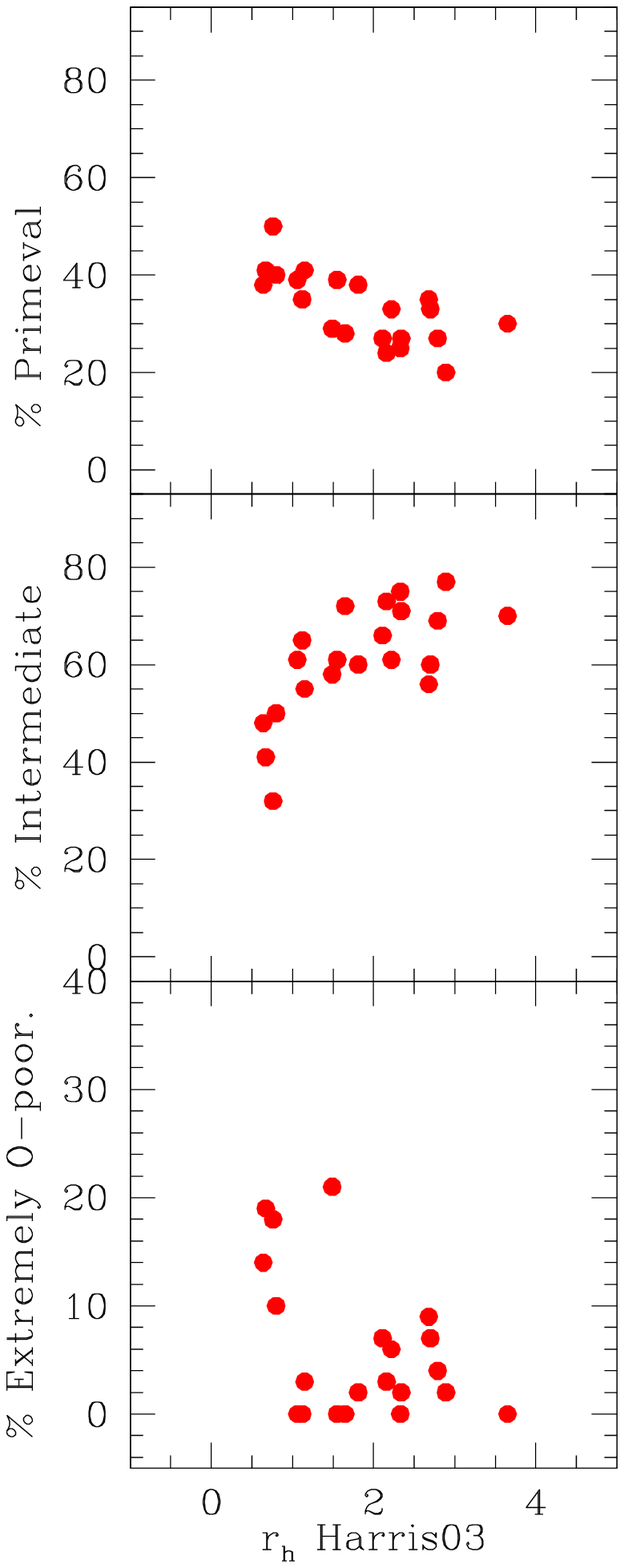}
\caption{Left: the fraction
of stars in the P, I and E components of second-generation stars as a function
of the total absolute magnitude (hence, mass) of clusters. Right: the same, 
as a function of the half-mass radius of the clusters.}
\end{figure}

The relations (also shown in Fig.~7) with half-mass radius ($r_h$) are very
tight, indicating that the shape of the anticorrelation was inprinted in the
early phases of the cluster life, since $r_h$ is one of the parameters less
sensitive to changes due to the cluster evolution, both internal and due to
external sources. More detailed studies, involving multivariate analysis, are 
under way.

At the moment, our preliminary findings can be summarized as follows:
\begin{itemize}
\item the extent of the Na-O anticorrelation is strictly related only to the
very blue and hot extreme of the HB
\item long anticorrelations are seen in clusters with both high mass $and$
favourable orbital parameters, taking them for long periods away from the
central regions of the Galaxy
\item in all GCs there is a surviving population of stars of primordial
composition, but
\item the bulk of the stellar population in GCs is composed of stars with
intermediate degrees of pollution
\item finally, an extremely O-poor component is observed only in some clusters
and at the moment is not entirely clear why.
\end{itemize}

In the near future we will present the full analysis of the clusters not already
published\footnote{We published results on: 
NGC~2808 (Carretta et al. 2006), NGC~6441 (Gratton et al. 2006, 2007),
NGC~6752 (Carretta et al. 2007a),
NGC~6218 (Carretta et al. 2007b), NGC~6838 (Carretta et al. 2007c).
}, a systematic study of the properties of second generation stars and 
their relations with global cluster parameters and the relation between the Na-O
anticorrelation and HB morphology (Carretta et al. 2009a,b and Gratton et al.
2009, in preparation, respectively).

\acknowledgments
This work is based on the data collected at ESO telescopes under the programmes
072.D-0507 and 073.D-0211.
We acknowledge partial financial support from PRIN-INAF 2005 
``Experimenting stellar nucleosynthesis in clean environments"

\end{document}